# Energy Efficient Tri-State CNFET Ternary Logic Gates

Sepher Tabrizchi, Fazel Sharifi, and Abdel-Hameed A. Badawy

*Abstract*—Traditional silicon binary circuits continue to face challenges such as high leakage power dissipation and large area of interconnections. Multiple-Valued Logic (MVL) and nano-devices are two feasible solutions to overcome these problems. In this paper, a novel method is presented to design ternary logic circuits based on Carbon Nanotube Field Effect Transistors (CNFETs). The proposed designs use the unique properties of CNFETs, for example, adjusting the Carbon Nanontube (CNT) diameters to have the desired threshold voltage and have the same mobility of P-FET and N-FET transistors. Each of our designed logic circuits implements a logic function and its complementary via a control signal. Also, these circuits have a high impedance state which saves power while the circuits are not in use. In an effort to show a more detailed application of our approach, we design a 2-digit adder-subtractor circuit. We simulate the proposed ternary circuits using HSPICE via standard 32nm CNFET technology. The simulation results indicate the correct operation of the designs under different process, voltage and temperature (PVT) variations. Moreover, a power efficient ternary logic ALU has been design based on the proposed gates.

## I. INTRODUCTION

Conventional silicon binary computing faces significant problems in terms of power and performance. Some of the most important challenges are the severe short channel effects of the Si-MOSFET and the restriction in the number of wires and pins of the chips that play more important roles than the device geometry. To overcome these challenges, one solution is to utilize non-silicon and non-binary circuits [1].

In order to use non-binary computing, the MVL paradigm has been introduced as an alternative to binary computing. In MVL, more than two logic values are used for data representation. More information can be conveyed over the same line and more data can be stored per memory cell by utilizing MVL techniques [1]. Also, using more than two significant logic levels leads to fewer computational steps, potentially fewer gates and considerable reduction in the number of interconnections and pins [2-4]. It was proven that e base (e ≈ 2.718) leads to the most efficient implementation of the switching systems among all MVL systems [2]. Therefore, ternary logic is superior to binary logic since three is the closest integer to e. Ternary logic provides the most efficiency with its lower energy consumption, as a result of the reduction in the number of interconnection wires and the cost of data movement.

In nano-scale CMOS devices leakage power is an important part of its total energy consumption. Other critical challenges are the reduced gate control and velocity saturation [5]. Therefore, to continue the historical improvement in chip transistor count, density and performance while operating at low-power, some emerging devices and technologies have attracted considerable attention in the recent years as alternatives for CMOS, such as quantum dot cellular automata (QCA), carbon nanotube field effect transistor (CNFET), single electron transistor (SET), nano magnetic devices, *etc.* [6-8]. Among these new technologies, CNFETs have attracted a lot of attention as a potential successor for CMOS because of its outstanding characteristics such as similarities with MOSFET, high carrier mobility, high $I_{ON}/I_{OFF}$ ratio, unique one dimensional band structure and near ballistic transportation [9, 10].

CNFET transistors are even more interesting, when they are used in designing MVL circuits. MVL circuit design is based on multiple threshold design techniques and adjusting the threshold voltage of CNFETs is easily possible by changing the diameter of the nanotubes [11, 12]. In recent years, some MOSFET and CNFET MVL circuits, have been presented for ternary and quaternary logic [10, 11, 13-21]. However, they have some critical drawbacks such as using very large ohmic resistors [13, 14], requiring obsolete depletion-mode MOSFET [15, 17-20], non-full swing nodes and limited fan-out. In this paper, we propose a novel method for designing ternary logic gates Buffer/NOT, AND/NAND, and OR/NOR. Each of the designs produces a logic function with its complimentary by a control signal. Moreover, a third state of high impedance is introduced to achieve power efficiency if none of the two

Sepehr. Tabrizchi is with the School of Computer Science Institute for Research in Fundamental Sciences (IPM), Tehran, Iran (s.tabrizchi@ipm.ir). Fazel. Sharifi, is with Department of Electrical and Computer Engineering, Graduate University of Advanced Technology, Kerman, Iran (f.sharifi@kgut.ac.ir). Abdel-Hameed. Badawy is with the Klipsch School of Electrical and computer Engineering, New Mexico State University, Las Cruces, NM. USA, (badawy@nmsu.edu).

possible gates is needed.

This paper extends the contributions of [22], in which ternary basic gates based on CNTFETs were presented. In this paper, we make additional contributions by presenting a two digit adder/subtractor as an application for the proposed basic gates in addition to detailed analysis with several figures of merit, such as propagation delay, power dissipation and the power-delay product (PDP). In addition, a low power ternary arithmetic logic unit (ALU) based on the presented circuits is designed and analyzed.

The rest of this article is organized as follows: Section II briefly reviews some background on CNFET devices and ternary logic. Section III describes the proposed designs. Section IV presents the simulation results and analyses. Finally, section V concludes the article.

## II. BACKGROUND

### A. Carbon Nanotube Field Effect Transistor

A carbon nanotube (CNT) is a sheet of graphene rolled up along a chirality vector [23]. The chirality vector of a CNT is defined by (n, m) pair. If n-m = 3k (k ϵ Z) then the CNT behaves like a metal, otherwise like a semiconductor [12].

Metallic nanotubes are attractive as future interconnects because of their superior properties, such as large current carrying capacity, and high thermal conductivity [24]. Also semiconducting nanotubes have great advantages. They can be used as channels in field effect transistors. They have high charge carrier mobility, lower sub-threshold swing and fewer parasitic elements [3]. Moreover, they are very attractive to Si semiconductor industry for the following reasons: The operation principle and the device structure are similar to CMOS devices; therefore, we can reuse the CMOS fabrication process and established CMOS design infrastructure. Also CNFETs show significant improvements in device performance metrics such as delay and power consumption in experimental results [25].

This three (or four) terminal device (CNFET) is turned on or off electrostatically via the gate and its threshold voltage ($V_{th}$). One of the most effective properties of CNFET, which makes it very suitable for designing digital circuits, is that the desired threshold voltage can be obtained by adopting proper diameter for the CNTs. The threshold voltage of a CNFET is given by the following equations, where, $e$ is the unit electron charge, $E_{bg}$ is the CNT bandgap, $a_0$ ($\approx$ 0.142 nm) is the carbon to carbon bond length in a CNT and $V_\pi$ ($\approx$ 3.033 eV) is the carbon π-π bond energy in the tight bonding model [26].

$$V_{th} \approx \frac{E_{bg}}{2e} = \frac{a_0 E_\pi}{eD_{CNT}} \approx \frac{0.436}{D_{CNT}(nm)} \quad (1)$$

$$D_{CNT} = \frac{\sqrt{3}a_0\sqrt{n^2 + m^2 + nm}}{\pi} \approx 0.0783\sqrt{n^2 + m^2 + nm} \quad (2)$$

According to Equation (2), the threshold voltage of a CNFET is inversely related to its CNT diameter.

Although CNFETs are promising, there are several challenges that need to be addressed. There are some difficulties for synthesis or growth of nanotubes with identical diameters and chiralities. Changes in tubes` diameter and wrapping angle, defined by the chirality indices (n, m), will shift the electrical conductivity and CNFET threshold voltage. However, many effective and feasible solutions have already been presented in the literature for growing CNTs with a specific chirality and setting the desired threshold voltage. Moreover, it is difficult to control the exact placement and alignment of CNTs at a VLSI scale. Mispositioned CNTs may cause incorrect logic functionality [27, 28].

### B. Ternary Logic and related works

Ternary logic consists of three significant logic levels represented by "0", "1" and "2" symbols. These logic levels are commonly counterpart to 0 V, ½$V_{DD}$ and $V_{DD}$ voltage levels, respectively. The ternary basic logic operations, which are the building blocks of many other complex logical and arithmetic quaternary circuits, can be defined according to (3), (4) and (5).

$X_i, X_j \in \{0, 1, 2\}$

$$X_i + X_j = Max\{X_i, X_j\} \quad (3)$$
$$X_i \cdot X_j = Min\{X_i, X_j\} \quad (4)$$
$$\overline{X_i} = 2 - X_i \quad (5)$$

where, $-$ denotes the arithmetic subtraction, the operations +, •, and $\overline{\phantom{x}}$ are the OR, AND, and NOT in ternary logic, respectively [28].

Three different type of ternary gates can be designed for each function. As an example for ternary inverter three logic gates can be defined; Standard Ternary Inverter (STI), Positive Ternary Inverter (PTI), and Negative Ternary Inverter (NTI). The truth tables of these gates for ternary inverter are shown in table II.

TABLE I. TRUTH TABLE OF PTI, NTI AND STI

| a | PTI(a) | NTI(a) | STI(a) |
|---|--------|--------|--------|
| 0 | 2      | 2      | 2      |
| 1 | 2      | 0      | 1      |
| 2 | 0      | 0      | 0      |

Some state-of-the-art CNFET-based ternary circuits have been presented in the literature. In [1], a CNFET-based ternary with large resistive loads, which are hard to implement and integrate with CNFETs and also cause performance degradation and wastes large area. Another ternary design has been presented in [12]. In [12], the resistors used in the design [1] is replaced with P-CNFET active loads which leads to less area overhead, larger noise margins and higher performance compared to the previous design.

The authors of [30] presented ternary logic circuits based on the complementary CNFET design style which uses three different threshold voltages. This design produce NTI, PTI and STI by a single circuit unlike previous designs. We adopt the design style of [30] throughout the designs of this paper.

## III. PROPOSED DESIGN(S)

In this section, ternary logic gates, including ternary Buffer/NOT, ternary AND/NAND and ternary OR/NOR gates, are introduced. The proposed designs use just two different diameters for their CNFETs. Based on Equations (1) and (2), for the CNFETs with 0.783 and 1.487 nm diameters, the chiral numbers are (10, 0) and (19,0), and the threshold voltages ($|V_{th}|$) are 0.557V and 0.293V, respectively. Moreover, in these designs a high impedance state can be produced, which can be used if any of the logic functions is not needed. This state consumes very lower power compared to the two other states which have static power dissipation. In summary, in this paper we propose a ternary family of logic circuits which can have three output states specified via a control signal.

### A. Ternary Buffer/Inverter

The proposed ternary Buffer/NOT circuit is shown in Fig.1. This circuit can act as a buffer or an inverter for a ternary input using a control signal (S). When the signal S = 0, the circuit acts as a ternary buffer. In this case, if IN = 0, both NTI and PTI nodes (shown in Fig. 1) will be $V_{DD}$, T5, T6 will be OFF and T3 and T4 will be ON, consequently, the output will be discharged to the ground through path 4. When IN = 2 ($V_{DD}$) the NTI and PTI nodes will be 0, and T1 and T2 will be ON and the output will be charged to $V_{DD}$ (path 3). When, IN = 1 (½$V_{DD}$), PTI and NTI are $V_{DD}$ and 0 respectively and T1, T2, T3 and T4 will be ON. So with a resistive voltage division, the output will be ½$V_{DD}$.

When S = 2 ($V_{DD}$), the circuit performs the NOT function. In this case, T1 and T2 are OFF and T5 and T6 are ON. Assume that IN=0, so STI and PTI will be $V_{DD}$ and consequently the output will be $V_{DD}$ through paths 1 and 2. For other inputs the output will be determined based on a resistive division.

Finally, when S=1 (½$V_{DD}$), the output will be high impedance (HZ). In this case, T1, T4, T5 and T6 will be OFF because of their threshold voltages which are higher than ½$V_{DD}$. Therefore, we do not have any path to the output. In this state, the circuit consumes very low power compared to the other states which consume static power. This state of the circuit is very useful in low power applications where no need for neither the buffer nor the inverter functionality continuously. The operation of this design is summarized in Table II.

Figures 2 and 3 show the voltage transfer characteristic (VTC) curves of the presented ternary buffer and inverter receptively. These schematics verify the correct operation and steep curve in the transition region, which leads to low average static power consumption. We used PTI(s) and NTI(s) for controlling T4 and T5 to have lower OFF current and consequently lower power consumption.

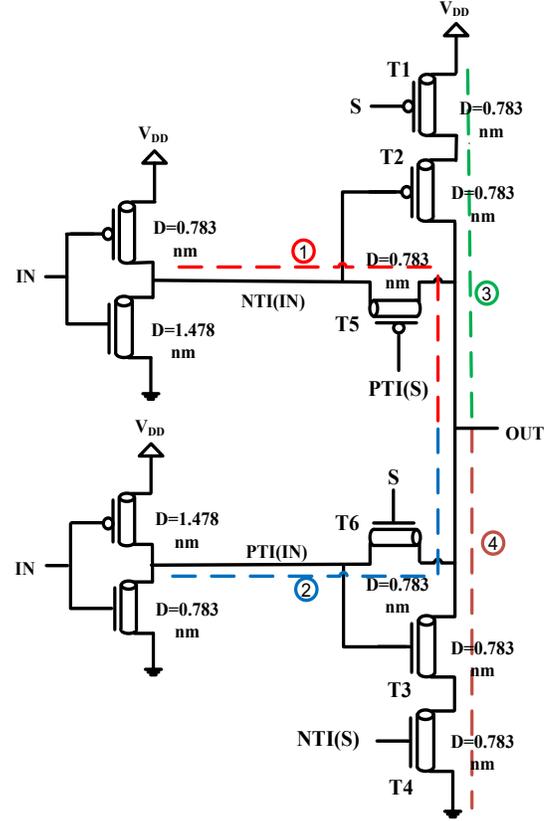

Fig. 1. The circuit design of our ternary Buffer/NOT Gate. This circuit has four different paths from $V_{DD}$ and ground to the output which are represented by numbers 1 -- 4. For each of the different inputs two or three paths will be active.

TABLE II. TRUTH TABLE FOR THE OPERATION OF THE BUFFER/NOT GATE

| S | IN | Out |
|---|---|---|
| 0 | 0 | 0 |
| 0 | 1 | 1 |
| 0 | 2 | 2 |
| 1 | 0 | HZ |
| 1 | 1 | HZ |
| 1 | 2 | HZ |
| 2 | 0 | 2 |
| 2 | 1 | 1 |
| 2 | 2 | 0 |

### B. Ternary AND/NAND – OR/NOR

Using the same design methodology for the ternary Buffer/NOT circuit, a new ternary AND/NAND and a new ternary OR/NOR are also designed. The operation principles of these two circuits are similar to the ternary Buffer/NOT gate. The schematics of the proposed AND/NAND and OR/NOR circuits are shown in Figs. 4 and 5, respectively. The operation of the ternary AND/NAND can be summarized as follows:

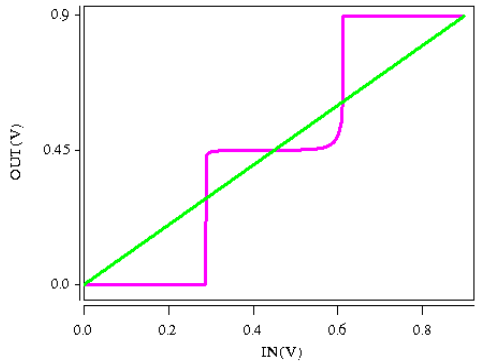

Fig. 2. The Voltage Transfer Characteristic (VTC) of our proposed ternary Buffer

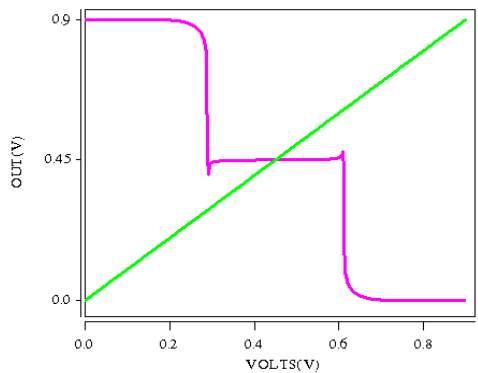

Fig. 3. The Voltage Transfer Characteristic (VTC) of our proposed ternary Inverter

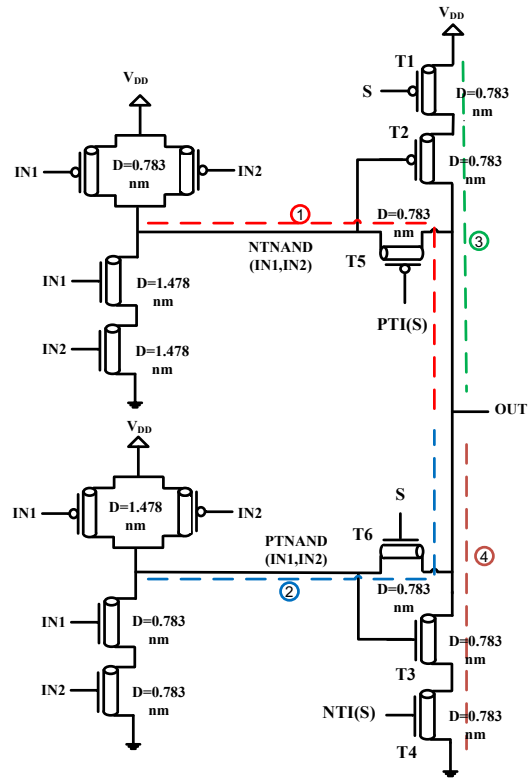

Fig. 4. The circuit design of our proposed ternary AND/NAND. If S = 0, this circuit acts as an AND gate. If S = 2, it acts as a NAND gate, and if S = 1, the output is High Impedance (HZ).

When both inputs (IN1, IN2) are around $V_{DD}$ and S = 0, NTNAND and PTNAND nodes are discharged to ground, so the output will be $V_{DD}$ through path 3. While one of the inputs is around $½V_{DD}$ and the other one is equal to or greater than $½V_{DD}$, both paths 3 and 4 are activated and the output will be $½V_{DD}$. Moreover when one or both of the inputs is around 0, both T3 and T4 are ON and the other paths to the output are disconnected, consequently the output is 0. In case of S = 2, T1 and T4 are OFF and the circuit implements the NAND functionality. During these operating conditions, the output will be determined based on paths 1 and 2. Finally, if S = 1 ($½V_{DD}$), all paths through the output will be disconnected and the output is HZ. The principle operation of the proposed ternary OR/NOR is very similar to the AND/NAND circuit operation.

The proposed circuits utilize CNFETs with only two different diameters for their CNTs, while most of the designs of ternary logic gates in the literature need at least three distinct diameters. This property improves robustness to process variation and enhances the manufacturability of the proposed circuits.

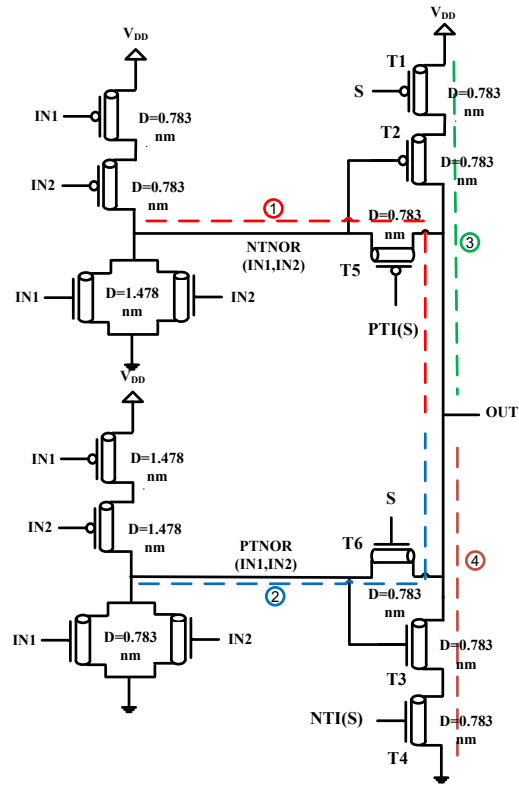

Fig. 5. The circuit design of our proposed ternary OR/NOR gate. If S = 0, this circuit acts as an OR gate. If S = 2, it acts as a NOR gate, and if S = 1, the output is High Impedance (HZ).

## C. Two Digit Ternary Adder/Subtractor

Multi digit ternary adder/subtractor has been designed using the proposed Buffer/Inverter. Two-digit adder/subtractor which is shown in Fig. 6, can performs addition and subtraction by a selector signal S. when S=0, the outputs of the binary buffer and the ternary Buffer/Inverter are zero and A respectively, so the circuit will add two digits. But, when S=2 ($V_{DD}$), the output of binary buffer and ternary Buffer/Inverter are 1 (½$V_{DD}$) and $\bar{A}$ respectively. So the circuit will perform the subtract operation. The applied binary buffer gets values 0 and 2 and produces 0 and 1 respectively. By using the proposed ternary Buffer/Inverter we could save $N$ multiplexer for $N$ digit adder/subtractor circuit.

## D. Ternary Arithmetic and Logic Unit (ALU)

In this section, two ternary arithmetic and logic units (ALUs) are presented. The proposed ALUs perform nine different logic and arithmetic operations. These functions are shown in Table III. The first design is illustrated in Fig. 7, which is based on multiplexers. The operations are controlled by two signals ($S_0$ and $S_1$) which are connected to the multiplexers selectors. When $S_0$ is 1, the ALU performs arithmetic operations (Addition, Subtraction and Increment), but when $S_0$ is 0 or 2, the ALU performs logic functions controlled by $S_1$ as shown in Table III. This design is simple and modular but uses multiplexers which increases the transistor count.

To decrease the number of transistors and take advantage of the proposed ternary logic gates, we present the second ALU design which is shown in Fig. 8. In this design, we have eliminated the multiplexers by using the third state (HZ) of the ternary gates proposed in the previous section.

Four customized circuits have been designed to produce additional control signals ($C_1$, $C_2$, $C_3$ and $C_4$) for ternary gates by using two main control signals ($S_0$ and $S_1$). These circuits are shown in Fig. 9 and the output of each circuit is in Table III.

The functionality of the proposed ternary ALU is briefly described in Table III. As it is indicated in this table, when $S_0$=1, the output of the logic unit is HZ and the ALU performs an arithmetic operation based on the value of $S_1$. But, when $S_0$=0 or $S_0$=2, the output of arithmetic unit will be HZ by signal $C_4$, and by adjusting a proper value for the $S_1$ signal a desired logic function will be performed. For example when $S_0$=0 and $S_1$=2, the $C_3$ output will be zero and $C_1$, $C_2$, $C_4$ are 1. Thus, the outputs of the two first logic gates (Buffer/Inverter and AND/NAND) and arithmetic circuit are HZ and consequently the ALU output will be the OR(A, B).

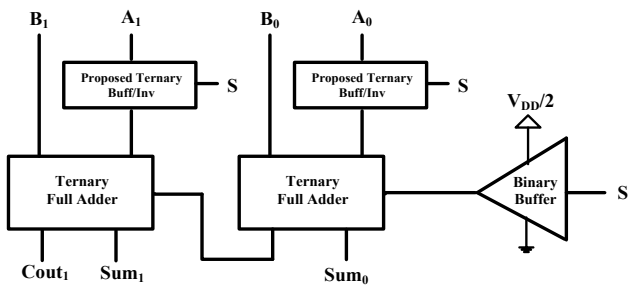

Fig. 6. A 2-digit ternary adder/subtractor design based on the proposed ternary Buffer/Inverter.

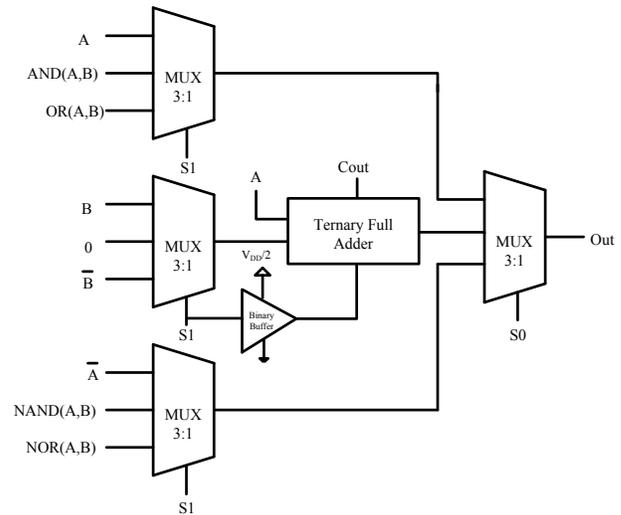

Fig. 7. Circuit schematic of the first ternary ALU.

TABLE III. THE TRUTH TABLE DETAILING THE OPERATIONS AND FUNCTIONALITY OF THE PRESENTED TERNARY ALU AND ITS CONTROL SIGNALS

| $S_0$ | $S_1$ | $C_1$ | $C_2$ | $C_3$ | $C_4$ | Logic Unit Output | Arith Unit Output | ALU Output |
|---|---|---|---|---|---|---|---|---|
| 1 | 0 | 1 | 1 | 1 | 0 | HZ | Add | Add |
| 1 | 1 | 1 | 1 | 1 | 0 | HZ | Increase | Increase |
| 1 | 2 | 1 | 1 | 1 | 0 | HZ | Subtract | Subtract |
| 0 | 0 | 0 | 1 | 1 | 1 | Buffer | HZ | Buffer |
| 0 | 1 | 1 | 0 | 1 | 1 | AND | HZ | AND |
| 0 | 2 | 1 | 1 | 0 | 1 | OR | HZ | OR |
| 2 | 0 | 2 | 1 | 1 | 1 | NOT | HZ | NOT |
| 2 | 1 | 1 | 2 | 1 | 1 | NAND | HZ | NAND |
| 2 | 2 | 1 | 1 | 2 | 1 | NOR | HZ | NOR |

The proposed ALUs can have more functions by adding more control signals. Also, having HZ state in the proposed ternary gates has two main advantages in the second design. (1) We do not need to use multiplexers in the ALU design which reduces both area and complexity. (2) Power efficiency can be reached by eliminating static power dissipation in unused ternary gates.

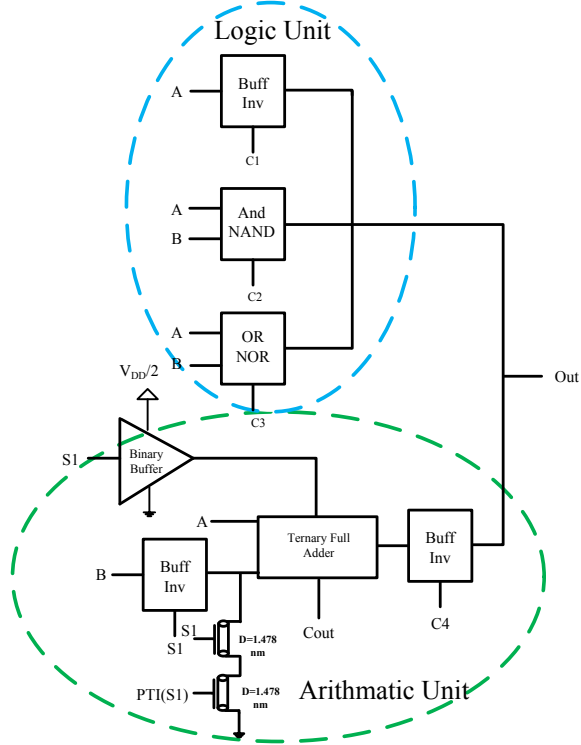

Fig. 8. Circuit schematic of the second ternary ALU using the presented ternary gates.

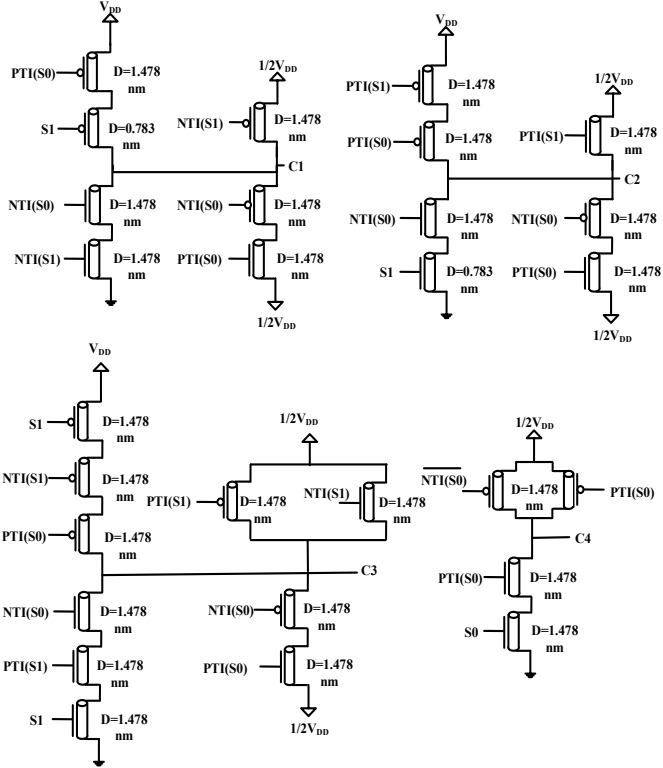

Fig. 9. The Customized circuits to produce control signals using in the presented ALU. These circuits generate C1, C2, C3 and C4 signals which act as control signals for ternary basic gates in ALU.

## II. SIMULATION RESULTS

In this section, the simulation results of the proposed circuits are presented. Simulations are conducted using the HSPICE simulator for 32 nm technology with the Stanford Compact SPICE model for CNFETs including the non-ideal and parasitic characteristics [31, 32]. Since this is the first attempt to design ternary logic gates with three output states, we could not compare our designs directly with state of the art designs. The output waveforms of the presented circuits are shown in Fig. 10, which confirms the correct operation of the designs.

Table IV provides the simulation results of the ternary designs including delay, average power consumption and power delay product (PDP). As indicated in Table III, the Buffer/NOT gate has the lowest delay. But due to the higher power consumption, it has a higher PDP compared to AND/NAND and OR/NOR designs. in order to have a fair comparison with previous ternary logic gates, we have simulated 2-digit ternary adder/subtractor using the proposed designs and previous designs presented in [11], and the results are shown in Table V. Based on the results using the proposed designs in a 2-digit adder/subtractor could save power consumption more than 12 times. Also the PDP of the circuit using the proposed designs are about 5 times better than the circuit using designs [11].

Moreover, the proposed ternary circuits are examined under different conditions and variations. The circuits are simulated with different temperatures from $0^oC$ -- $100^oC$. As it is clear from Fig. 11, these designs have almost constant PDP variation for all temperatures due to the high thermal stability of CNFETs.

Figure 12 shows the PDP variation of the circuits under different supply voltages. Designs are simulated in 0.8, 0.9 and 1V and the proposed ternary NAND has lower PDP because of its lower power consumption under all supply voltages.

The operation of the ternary gates is also examined in the presence of process variation. One of the most important challenges in nanoscale devices is sensitivity to process variation, which can negatively impact the robustness of the circuits. It has been proven experimentally that the dominant source of variation in CNFET circuits is the nanotube density variations, which mainly results from variations in the spacing between CNTs on the substrate (*pitch*) and variations in the surviving CNT count after metallic CNT removal techniques [31-33]. Therefore, we used a Monte Carlo simulation to evaluate the CNT density variation with up to ±15% Gaussian distributions and variation at the ± $3\sigma$ levels. As in Fig. 13, all the designs are robust against CNT density variation.

In Table VI, delay, power and PDP of the ternary ALUs have been presented. In this table logic and arithmetic units delay are presented separately. As it was predicted, delay and power consumption of the first ALU are slightly more than the second ALU. The PDP of the first ALU is about 19% more than the second one.

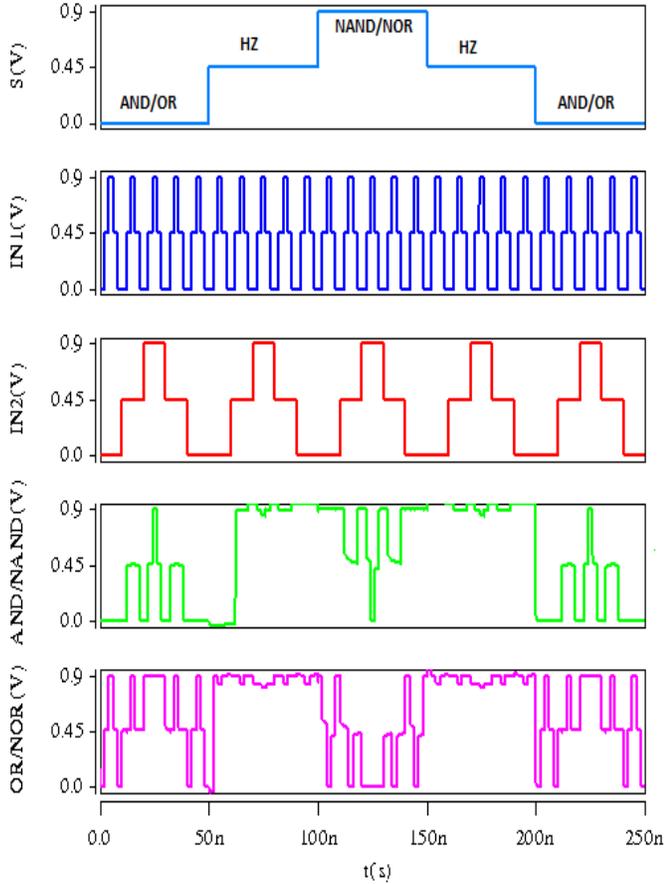

Fig. 10. Output waveforms of the proposed ternary AND/NAND and OR/NOR gates. When S = 0, the circuits perform AND/OR functions. When S = 1, the output is high impedance. When S = 2, the circuits perform NAND/NOR operation.

TABLE IV. DELAY, POWER AND POWER DELAY PRODUCT (PDP) OF OUR DESIGNED CIRCUITS

| Design | Delay (E-11s) | Maximum Delay (E-11s) | Power (E-7W) | PDP (E-17J) |
|---|---|---|---|---|
| Buffer | 1.878 | 1.878 | 17.711 | 3.326 |
| NOT | 0.781 | | | |
| AND | 2.762 | 2.762 | 7.0485 | 1.946 |
| NAND | 1.202 | | | |
| OR | 2.732 | 2.732 | 10.54 | 2.879 |
| NOR | 1.314 | | | |

TABLE IV. SIMULATION RESULTS OF 2-DIGIT TERNARY ADD/SUB USING THE PROPOSED DESIGNS AND DESIGNS [11]

| Design | Maximum Delay (E-11s) | Power (E-7W) | PDP (E-17J) |
|---|---|---|---|
| 2 Digit Add/Sub Using proposed designs | 13.80 | 47.57 | 65.64 |
| 2 Digit Add/Sub Using designs [11] | 5.79 | 617.2 | 357.3 |

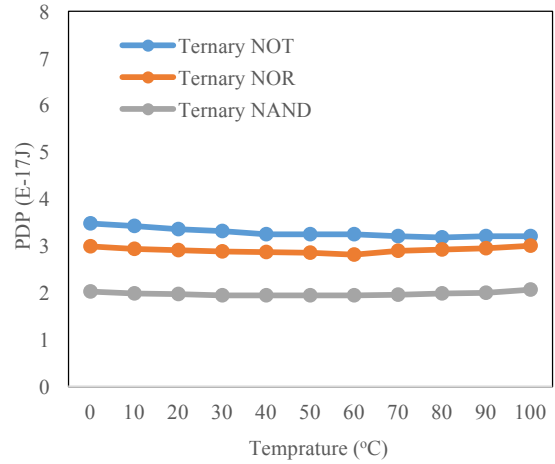

Fig. 11. PDP variation against temperatures variation

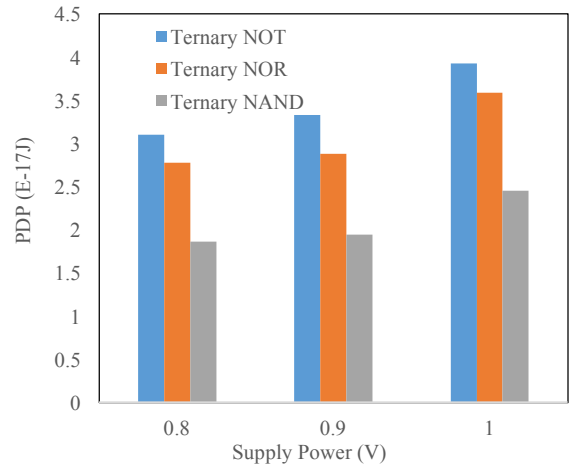

Fig. 12. PDP variation in different supply voltages

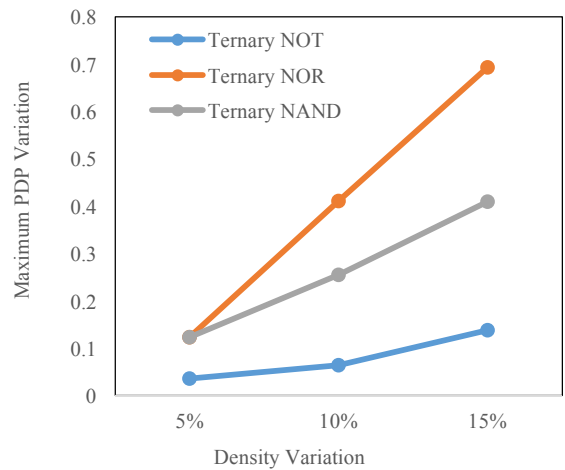

Fig. 13. Maximum PDP variation against CNTs density variation. Density includes the CNT's pitch and the number of CNTs under gate.

TABLE VI. DELAY, POWER AND POWER DELAY PRODUCT (PDP) OF TERNARY ALUS

| Design | Logic Delay (E-11s) | Arith Delay (E-11s) | Power (E-6W) | PDP (E-17J) |
|---|---|---|---|---|
| 1st ALU | 2.175 | 6.699 | 13.11 | 87.82 |
| 2nd ALU | 1.652 | 6.307 | 11.68 | 73.66 |

## III. CONCLUSION

In this paper, three output-states ternary logic circuits are presented. Each of the presented circuits can perform a logic function or its complement via a control signal. When the circuits are idle *i.e.* not in use, the output is high impedance (HZ), which lowers the power consumption. We design these circuits using carbon nano-tube field effect transistors (CNFETs). CNFETs are very appropriate for designing MVL circuits because of their ability to set the desired threshold voltage by adjusting the tubes' diameters. Moreover, two-digit adder/subtractor and two ternary ALUs have been designed using the proposed gates. The second ALU can reach the power efficiency by using the high impedance (HZ) state of the gates when they are not in use. Circuits are simulated using HSPICE simulator with 32nm CNFET technology under different conditions. The results show robustness under process variation, temperature, and supply voltage. The AND/NAND gate has the lowest PDP compared to the other proposed designs because of its lower power dissipation. It has almost 48% lower PDP compared to the OR/NOR gate.

## ACKNOWLEGEMENT

The authors would like to thank the reviewers of their NANO'17 paper for their feedback which helped influence the presentation of this article. The authors would like to thank the members of the PEARL laboratory at New Mexico State University for the discussions and help.